\documentclass[aps,pra,twocolumn,preprintnumbers,nofootinbib,amsmath,amssymb,footinbib]{revtex4}
\usepackage{mathtools}

\DeclareMathAlphabet{\mathcalligra}{T1}{calligra}{m}{n}
\DeclareMathAlphabet{\mathpzc}{OT1}{pzc}{m}{it}

\usepackage{graphicx,epsfig}
\usepackage{bm}
\usepackage{dcolumn}
\usepackage{xcolor}
\usepackage[breaklinks=true,colorlinks,citecolor=blue,linkcolor=blue,urlcolor=blue]{hyperref}

\def\sc#1#2#3{Science {\bf #1}, #2 (#3)}
\def\prv#1#2#3{Phys. Rev. {\bf #1}, #2 (#3)}
\def\rmp#1#2#3{Rev. Mod. Phys. {\bf #1}, #2 (#3)}
\def\prl#1#2#3{Phys. Rev. Lett. {\bf #1}, #2 (#3)}
\def\pra#1#2#3{Phys. Rev. A {\bf #1}, #2 (#3)}
\def\prb#1#2#3{Phys. Rev. B {\bf #1}, #2 (#3)}

\def\pre#1#2#3{Phys. Rev. E {\bf #1}, #2 (#3)}

\def\epjd#1#2#3{Eur. Phys. J. D {\bf #1}, #2 (#3)}

\def\ejp#1#2#3{Eur. J. Phys. {\bf #1}, #2 (#3)}

\def\noi{\noindent}
\def\bc{\begin{center}}
\def\ec{\end{center}}
\newcommand{\bea}{\begin{equation}}
\newcommand{\eea}{\end{equation}\noi}
\newcommand{\ber}{\begin{eqnarray}}
\newcommand{\eer}{\end{eqnarray}\noi}
\begin{document}
\title{Opto-criticality in Drude metals near the plasma frequency}

\author{Bikram Keshari Behera$^1$}
\author{Rhitabrata Bhattacharyya$^2$}
\author{Shyamal Biswas$^1$}\email{sbsp@uohyd.ac.in}

\affiliation{$^1$School of Physics, University of Hyderabad, C.R. Rao Road, Gachibowli, Hyderabad-500046, India\\
$^2$Central GST \& Central Excise, Kolkata South Commissionerate, GST Bhawan, Kolkata-700107, India
}

\date{\today}

\begin{abstract}
Within a simple unified framework of classical electrodynamics that incorporates background bound charges and currents, we have analytically determined the attenuation constant, phase constant, and reflectivity of Drude metals across the full frequency spectrum ($0 < \omega < \infty$). Our result for reflectivity agrees well with existing experimental data for both plasmas and seawater. Notably, for highly conductive Drude metals ($\omega_p \tau \gg 1$), the attenuation constant simplifies to $k_- \simeq \sqrt{\frac{\mu \epsilon}{2}} \sqrt{\omega_p^2 - \omega^2 + \vert{}\omega_p^2 - \omega^2\vert{}}$ across a broad high-frequency range spanning the plasma frequency $\omega_p$. The branch-point singularity at $\omega=\omega_p$ in this relation reveals opto-critical behavior near $\omega=\omega_p$ in properties such as the attenuation constant, group velocity, and complex dielectric constant, for which we have determined the corresponding opto-critical exponents. Furthermore, we have incorporated quantum corrections into these optical properties via the Drude–Sommerfeld model with Thomas-Fermi screening.
\end{abstract}


\maketitle    

\section{Introduction}
The skin depth ($\delta=1/k_{-}$, the inverse of the attenuation constant), is the characteristic length scale for electromagnetic wave penetration in a conductor \cite{Griffiths}. Along with the phase constant and reflectivity, it governs the frequency-dependent optical properties of conducting media \cite{Griffiths, Born}. Studying these response functions in conductors \cite{Roberts} and solids \cite{Linz, Wooten, Fox, Yarmohammadi, Murakami} drives advances in semiconductor and condensed matter physics \cite{Cardona, Devereaux, Blaber, Guistino, Russo, Vries, Arafune, Tonauer, Rahaman, Tiwari, Huang}, as well as ocean science applications such as submarine radio communication \cite{Moore, Smolyaninov, Talley, Adcock, Jarman}.

An electrical conductor is often treated as a linear medium with permittivity $\epsilon>\epsilon_0$ and permeability $\mu>\mu_0$, reflecting bound charges and currents \cite{Griffiths, Born, Rothwell}. It also contains conduction electrons that follow Ohm's law, yielding a complex conductivity $\sigma=\sigma_0/[1-i\omega\tau]$ under the classical Drude model for a real relaxation time $\tau$ when exposed to an electromagnetic wave of angular frequency $\omega$ \cite{Drude, Wooten, Fox, Jackson, Zangwill}. We refer to such a medium as a Drude metal. While ordinary metals like copper, silver, and aluminum act as good conductors, seawater is a poor conductor with a strongly frequency-dependent response \cite{Griffiths}. Thus, its reflectivity and transmissivity are key parameters for marine electromagnetic propagation \cite{Moore, Smolyaninov}.

Electromagnetic wave propagation in Drude metals is well documented across two primary regimes: (i) quasi-static ($\omega\tau \ll 1$): classically modeled via Maxwell's equations using real conductivity in a linear medium with background bound charges and currents ($\epsilon > \epsilon_0$, $\mu > \mu_0$) \cite{Biswas-2026, Griffiths, Rothwell}, and (ii) high-frequency ($\omega\tau \gg 1$): classically modeled using complex conductivity in the absence of background bound charges and currents ($\epsilon = \epsilon_0$, $\mu = \mu_0$) via the Lorentz oscillator model in the zero-resonance limit ($\omega_0 \to 0$) \cite{Wooten, Fox, Jackson, Zangwill, Bronin}. Both the regimes were previously unified under this Lorentz framework with $\omega_0 \to 0$ (as well as, under the Drude model), though omitting background bound charges and currents, to analyze phase and group velocities in conductors \cite{Oughstun, Zangwill}. However, a unification incorporating background bound charges and currents remains unaddressed—even in quantum materials \cite{Boschini}—despite being relatively straightforward. Meanwhile, high carrier concentrations enable the study of critical optical behavior near the plasma frequency, analogous to the ferromagnetic–paramagnetic phase transition near a classical critical point \cite{WilsonNL}. To date, no such study has been reported. Furthermore, quantum many-body effects on this criticality, alongside the influence of background charges and currents in the frequency domain, are highly relevant.

In this article, we present a relatively straightforward unified framework based on Maxwell's equations and the complex Drude model to analyze the full frequency response ($0 < \omega\tau < \infty$) of conducting media, explicitly retaining background bound charges and currents. Rather than introducing a new microscopic model, this approach provides a common foundation to determine the complex dielectric function, attenuation, phase constant, group velocity, and reflectivity across both low- and high-frequency limits, as wall as across low- and high-carrier density limits. The formulation is bench-marked against experimental reflectivity data for media with contrasting carrier densities: highly conductive dense air and hydrogen Z-pinch plasmas \cite{Kurilenkov}, and weakly conductive seawater \cite{Saxton} with strong bound-charge contributions. In high-carrier-density regimes ($\omega_p\tau \gg 1$), where some of the above mentioned optical properties vary sharply around the plasma frequency $\omega_p$, the model reveals scaling laws governing optical criticality (\textit{hence-forth opto-criticality}) in the frequency domain. We further incorporate the Drude–Sommerfeld model \cite{Sommerfeld} and Thomas–Fermi screening \cite{Thomas-Fermi} to account for low-temperature quantum-statistical effects on optical properties, particularly their effect on modifying plasma frequency.

The calculation in this article begins with an evanescent plane-wave solution to Maxwell’s equations for a Drude metal with $\epsilon > \epsilon_0$, $\mu > \mu_0$, and a complex AC conductivity. We analytically derive the attenuation constant, phase constant, and reflectivity across all frequencies ($0 < \omega < \infty$), unifying the low- and high-frequency limits in a relatively straightforward manner. We evaluate wave propagation at high carrier concentrations in both quasi-static and non-quasi-static regimes, extracting optical critical (henceforth \textit{opto-critical}) exponents near the plasma frequency for the attenuation constant, phase constant, group velocity, and complex dielectric constant. Finally, we incorporate low-temperature ($T \ll T_F$) quantum many-body effects via the Drude–Sommerfeld model \cite{Sommerfeld} combined with Thomas–Fermi screening \cite{Thomas-Fermi}.
  
\section{Attenuation constant and other optical properties of Drude metals with background bound charges and currents}
Let us consider a plane electromagnetic wave $\vec{E}(\vec{r},t)=\vec{E}_0 e^{i(\vec{k}_0\cdot\vec{r}-\omega t)}$ (or $\vec{B}(\vec{r},t)$) with $k_0=\omega/c$\footnote{Here, $\vec{E}$ is the electric field, $\vec{B}$ is the magnetic field, $k_0$ is the incident wave-number, $\omega$ is the angular frequency, and $c=1/\sqrt{\mu_0\epsilon_0}\simeq3\times10^8\text{ m/s}$ is the speed of light in vacuum.}, normally incident from vacuum onto a Drude metal \cite{Griffiths, Jackson}. Inside the conductor, Ohm's law ($\vec{J}_f=\sigma\vec{E}$) leads to the driven wave equations \cite{Griffiths}
\begin{equation}\label{eq1}\nabla^2 \vec{E}=\mu\sigma\frac{\partial \vec{E}}{\partial t}+\mu\epsilon\frac{\partial^2\vec{E}}{\partial t^2}~~\&~~\nabla^2 \vec{B}=\mu\sigma\frac{\partial \vec{B}}{\partial t}+\mu\epsilon\frac{\partial^2\vec{B}}{\partial t^2},
\end{equation}
valid at long time scales ($t\gg\epsilon/\sigma_0$) under continuous excitation at frequency $\omega$. Background bound charges and currents set the linear medium parameters ($\epsilon>\epsilon_0$ and $\mu>\mu_0$), while finite DC resistivity ($1/\sigma_0$) causes wave attenuation. At normal incidence, the evanescent plane-wave solutions to Eqns.~(\ref{eq1}) inside the conductor are
\begin{equation}\label{eq3}
\vec{E}(\vec{r},t)=\vec{E}_0 e^{i(\vec{k}\cdot\vec{r}-\omega t)}~~\&~~\vec{B}(\vec{r},t)=\vec{B}_0 e^{i(\vec{k}\cdot\vec{r}-\omega t)},
\end{equation}
which retain the incident frequency $\omega$ but acquire a complex wave vector $\vec{k}=(k++ik-)\hat{k}$ (with $k_+>0$ and $k_->0$) due to the AC Drude conductivity \cite{Wooten,Fox,Jackson,Zangwill}
\begin{equation}\label{eq5}
\sigma=\frac{\sigma_0}{1-i\omega\tau}.
\end{equation}
Here, $\sigma_0=n_e e^2\tau/\vert{}m_e^*\vert{}$\footnote{For p-type semiconductors, $-e$ and $n_e$ are replaced by the hole charge $e$ and concentration $n_p$, while $m_e^*$ is replaced by the hole effective mass $m_h^*$ (which can be negative in band theory). The absolute value $\vert{}m_e^*\vert{}$ ensures positive DC conductivity in both cases.} is the DC conductivity, $n_e$ is the conduction electron density, $-e$ is the electron charge, and $m_e^*$ is the effective electron mass in the lattice. The real part of $\sigma$ arises from inelastic collisions of free carriers with lattice impurities, whereas the imaginary part reflects field oscillations inside the medium. The free charges oscillate at the plasma frequency $\omega_p=\sqrt{\sigma_0/(\tau\epsilon)}$ \cite{Fox,Zangwill}. For $\omega\gtrsim\omega_p$, field oscillations dominate over collisional damping, making the optical response (including attenuation, phase velocity, and group velocity) particularly dynamic near $\omega=\omega_p$.

\subsection{Attenuation constant with background charges and currents for all frequencies}
While the real part of the complex wave number ($k=k_+ + i k_-$) in Eqn.~(\ref{eq3}) is known as the phase constant ($k_+$) and is responsible for the propagation of the electromagnetic wave inside a conductor, its imaginary part is the attenuation constant ($k_-$), which is responsible for attenuation on the length scale of the skin depth ($\delta=1/k_-$). Let us now determine the attenuation constant $k_-$ of a Drude metal over the entire frequency range ($0<\omega<\infty$). The evanescent plane-wave solutions [Eqn.~(\ref{eq3})] to the wave equations [Eqn.~(\ref{eq1})] with the complex wave vector $\vec{k}=(k_+ + i k_-)\hat{k}$ and the complex conductivity $\sigma$ [Eqn.~(\ref{eq5})] yield $k^2 = i\mu\sigma\omega + \mu\epsilon\omega^2$, which leads to
\begin{eqnarray}\label{eq9}
k_{\pm}&=&\sqrt{\frac{\mu\epsilon\omega^2}{2(1+\omega^2\tau^2)}}\bigg[\sqrt{(1+\omega^2\tau^2)\big[1+(\omega^2-\omega_p^2)^2\tau^2/\omega^2\big]}\nonumber\\&&
\pm\big[1+(\omega^2-\omega_p^2)\tau^2\big]\bigg]^{1/2},
\end{eqnarray}
where $\omega_p=\sqrt{\frac{n_e e^2}{\vert{}m_e^*\vert{}\epsilon}}$ \cite{Fox,Zangwill} is the plasma frequency of the conductor. Eqn.~(\ref{eq9}) represents the Maxwell–Drude response of the complex wave number and provides a unified result across the entire frequency range that accounts for the background bound charges ($\epsilon>\epsilon_0$) and currents ($\mu>\mu_0$). This unification, though relatively straightforward, allows us to study not only the low-frequency behavior of the conductor (especially where the effects of bound charges and currents are significant), but also its high-frequency behavior, both below and above the plasma frequency ($\omega_p$), for high carrier concentrations ($\omega_p\tau=\sqrt{n_e e^2 \tau^2 / \vert{}m_e^*\vert{}\epsilon} \gg 1$).

\subsection{Special cases of the attenuation constant for high carrier concentration}
\subsubsection{A case of high carrier concentration in the quasi-static regime}
The quality of an electrical conductor is determined by its skin depth ($\delta$) or attenuation constant ($k_- = 1/\delta$). A Drude metal acts as a good conductor (large skin depth) when the very low-frequency condition $\omega\tau \ll \omega_p^2\tau^2$ is met, or as a poor conductor (small skin depth) under low carrier concentrations at low frequencies ($\omega_p^2\tau^2 \ll \omega\tau \ll 1$). Thus, Eqn. (\ref{eq9}) yields:
\begin{equation}\label{eq10}
k_- \simeq \sqrt{\frac{\omega\sigma_0\mu}{2}}
\end{equation}for a good conductor \cite{Griffiths}, and
\begin{equation}\label{eq11}
k_- \simeq \frac{\sigma_0}{2}\sqrt{\frac{\mu}{\epsilon}}
\end{equation}for a poor conductor \cite{Griffiths}. Eqns. (\ref{eq10}) and (\ref{eq11}) align with standard literature results obtained for $\omega\tau \ll \omega_p^2\tau^2 \ll 1$ and $\omega_p^2\tau^2 \ll \omega\tau \ll 1$, respectively, where static conductivity ($\sigma \simeq \sigma_0$) is assumed in the quasi-static regime ($\omega\tau \ll 1$) \cite{Griffiths,Zangwill}. Our approach differs because it does not assume $\sigma \simeq \sigma_0$ to derive Eqn.~(\ref{eq10}). Notably, deriving Eqn.~(\ref{eq9}) does not require the additional constraint $\omega_p^2\tau^2 \ll 1$; it holds even for high carrier concentrations ($\omega_p^2\tau^2 \gg 1$), provided the low-frequency condition $\omega\tau \ll \omega_p^2\tau^2$ is satisfied. 

\begin{figure}
\includegraphics[height=6cm,width=8cm]{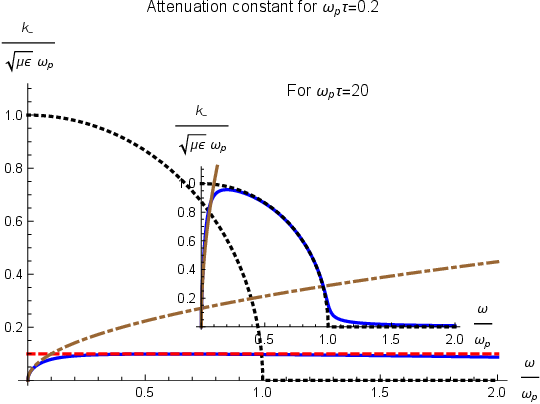}
\caption{Attenuation constant versus frequency for $\omega_p\tau=0.2$ (solid: Eqn.~\ref{eq9}, dashed: Eqn.~\ref{eq10}, dash-dotted: Eqn.~\ref{eq11}, dotted: Eqn.~\ref{eq12}). The inset shows the same for $\omega_p\tau=20$.	
}
\label{fig1}
\end{figure} 

\subsubsection{A case of high carrier concentration in the non-quasi-static regime: an opto-criticality at $\omega=\omega_p$}
The quality of a Drude metal with background bound charges and currents has not been analyzed for high carrier concentrations ($\omega_p\tau \gg 1$) in the non-quasi-static regime ($\omega\tau \gtrsim 1$), particularly near the plasma frequency ($\omega \sim \omega_p$). In this regime, the attenuation constant from Eqn.~\eqref{eq9} simplifies to:
\begin{equation}\label{eq12}
k_-\simeq\sqrt{\frac{\mu\epsilon}{2}}\sqrt{\omega_p^2-\omega^2+|\omega_p^2-\omega^2|}.
\end{equation}
It is interesting to note that Eqn.~\eqref{eq12} has a branch-point singularity at $\omega=\omega_{p}$. In the limit $\omega_p\tau \rightarrow \infty$, Eqn.~\eqref{eq12} yields $k_- = 0$ for $\omega \ge \omega_p$—a result previously established without background charges and currents. For $\omega < \omega_p$, however, Eqn.~\eqref{eq12} reduces to $k_- = \sqrt{\mu\epsilon}\sqrt{\omega_p^2 - \omega^2}$, which has not been reported in the literature. These distinct behaviors indicate opto-critical scaling near $\omega = \omega_p$:
\begin{equation}\label{eq13}
k_{-}\simeq\sqrt{2\mu\epsilon}\omega_p|1-\omega/\omega_p|^\nu,
\end{equation}
with an opto-critical exponent of $\nu = 1/2$ for $\omega < \omega_p$ and $\nu \rightarrow \infty$ for $\omega > \omega_p$, analogous to the ferromagnetic-paramagnetic phase transition near a critical temperature \cite{WilsonNL}. 

Figure~\ref{fig1} shows the frequency dependence of the attenuation constant [Eqn.~\eqref{eq9}] for low ($\omega_p\tau = 0.2$) and high ($\omega_p\tau = 20$, inset) carrier concentrations. The solid, dashed, dash-dotted, and dotted lines correspond to Eqns.~\eqref{eq9}, \eqref{eq10}, \eqref{eq11}, and \eqref{eq12}, respectively. The overlap of the solid and dash-dotted lines confirms that the condition for a good Drude metal is $\omega\tau \ll \omega_p^2\tau^2$ rather than $\omega\tau \ll \omega_p^2\tau^2 \ll 1$, which holds for both low and high carrier concentrations. At high concentration, the metal also behaves as a good conductor for $\omega \gtrapprox \omega_p$ \cite{Zangwill}. Conversely, the overlap of the solid and dashed lines shows that the condition for a poor Drude metal is $\omega_p^2\tau^2 \ll \omega\tau \ll 1$, which occurs only at low concentration \cite{Zangwill}. Interestingly, at high concentration, the metal becomes poor within $1/\tau \lesssim \omega \lnapprox \omega_p$. Finally, the dotted curve in the inset illustrates the transition in opto-critical exponent from $\nu = 1/2$ (for $\omega < \omega_p$) to $\nu \rightarrow \infty$ (for $\omega > \omega_p$) as $\omega_p\tau \rightarrow \infty$.

\subsection{The opto-criticality in the other optical properties at the high carrier concentration}

\begin{figure}
\includegraphics[height=6cm,width=8cm]{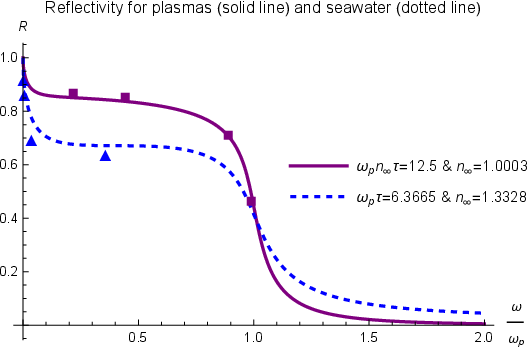}
\caption{The solid line follows Eqn.~(\ref{eq13a}) for $\omega_{p}n_\infty\tau=12.5$, $\mu=\mu_0$, and $n_\infty=1.0003$ (plasma), while the dotted line uses the same equation with $\omega_{p}\tau=6.3665$, $\mu=\mu_0$, and $n_\infty=1.3328$ (seawater). Squares and triangles denote adapted experimental data at normal incidence for dense Z-pinch air-hydrogen plasmas ($n_e\simeq10^{27}$~m$^{-3}$, $T\simeq5\times10^4$~K, $\Gamma\simeq0.2$) \cite{Kurilenkov,Skowronek} and seawater at $10\,^\circ\text{C}$ \cite{Saxton}, respectively.
}
\label{fig2}
\end{figure}

The unified results [Eqn.~(\ref{eq9})] for the phase constant ($k_{+}$) and attenuation constant ($k_{-}$) also unify other optical properties of the Drude metal—including skin depth ($\delta=1/k_-$), absorption coefficient ($\alpha=2k_{-}$), complex refractive index ($\tilde{n}=\frac{c}{\omega}k=\frac{c}{\omega}[k_++ik_-]$\footnote{Here, $c=1/\sqrt{\mu_0\epsilon_0}\simeq3\times10^8$~m/s is the speed of light in vacuum, $ck_+/\omega=n_\omega$ is the absolute refractive index, and $ck_-/\omega=\kappa_\omega$ is the extinction coefficient.}), complex dielectric constant ($\tilde{\epsilon}/\epsilon_0=\tilde{n}^2$), phase speed ($v_p=\omega/k_+$), group velocity ($\vec{v}_g=\frac{\partial\omega}{\partial k_{+}}\hat{k}$), reflectivity ($R$), and transmissivity ($1-R$) across the entire frequency range. Consequently, the absorption constant ($\alpha=2k_-$), attenuation constant ($k_-$), and related optical properties can be conveniently determined from experimental data for reflectivity \cite{Kurilenkov,Skowronek,Reinholz}, given by the Fresnel formula for normal incidence,
\begin{eqnarray}\label{eq13a}
R(\mu,\epsilon,\omega,\tau)=\bigg|\frac{\tilde{n}-1}{\tilde{n}+1}\bigg|^2=\frac{[k_+-\omega/c]^2+k_-^2}{[k_++\omega/c]^2+k_-^2},
\end{eqnarray}
and the absolute refractive index ($n_\omega=ck_+/\omega$\footnote{For mechanical refraction, see Ref.~\cite{Behera}.}) \cite{Verleur}, or vice versa. Although Eqn.~(\ref{eq13a}) appears in its standard form, the result is novel due to the new expressions for the phase and attenuation constants ($k_{\pm}$) [Eqn.~(\ref{eq9})] across all frequencies, accounting for bound charges and background currents in the conductor.

Figure~\ref{fig2} compares Eqn.~(\ref{eq13a}) with adapted experimental data of Refs. \cite{Kurilenkov,Skowronek,Saxton}. For dense Z-pinch air-hydrogen plasmas (good conductors), the experimental data \cite{Kurilenkov,Skowronek} show excellent agreement with Eqn.~(\ref{eq13a}) (solid line) using parameters\footnote{Here, $\tau=\frac{1}{0.08\omega_{p}n_\infty}$ is the relaxation time, $\omega_{p}=1.89\times10^{15}$~Hz ($\pm10$--$20\%$) is the plasma frequency \cite{Kurilenkov,Skowronek}, and $n_\infty\simeq1.000273\pm1$ \cite{Voronin} is the background air refractive index.} $\omega_{p}n_\infty\tau\simeq12.5$, $\mu\simeq\mu_0$, and $n_\infty\simeq1.0003$. Likewise, experimental data for seawater \cite{Saxton} (a relatively poor conductor) match reasonably well with Eqn.~(\ref{eq13a}) (dotted line) for parameters $\omega_{p}\tau\simeq6.3665$, $\mu\simeq\mu_0$, and $n_\infty\simeq1.3328$\footnote{Here, $\omega_{p}\simeq5.2616\times10^{11}$~Hz \cite{Saxton} is the ionic plasma frequency primarily from NaCl, $\tau=\frac{\epsilon_s\epsilon_0}{\sigma_0}=(12.1\pm0.1)\times10^{-12}$~s is the relaxation time, $\epsilon_s=72\pm1$ is the static dielectric constant at $10^\circ$C and $0.62\pm0.1$~N concentration \cite{Saxton}, and $n_\infty\simeq1.33282\pm0.000001$ \cite{Harvey} is the background pure water refractive index at $10^\circ$C and $0.1$~MPa for $\lambda=0.6328~\mu$m.}. For seawater, the dotted curve highlights a significant drop in reflectivity from unity for $\omega/\omega_{p}\ll1$, driven primarily by DC conductivity ($\sigma_0=n_ee^2\tau/\vert{}m_e^*\vert{}$) and the refractive-index-dependent term $-\frac{2n_\infty\sqrt{2n_\infty\omega}}{\sqrt{\tau}\omega_{p}}$. Conversely, as $\omega/\omega_{p}\rightarrow\infty$, the reflectivity tail approaches $\frac{[n_\infty-1]^2}{[n_\infty+1]^2}$, which depends solely on bound charge carriers (electrons). This limiting value provides a direct physical measure of their contribution. Furthermore, a cross-over, indicating an opto-criticality in reflectivity, is apparent near the plasma frequency ($\omega_p\tau\gg1$), corresponding to high carrier density in the plasma and high salinity in seawater.

Let us now focus on the opto-critical quantities—specifically the group velocity and complex dielectric constant near $\omega=\omega_p$ in the limit $\omega_p\tau\rightarrow\infty$. Corresponding to the attenuation constant $k_-$ in Eqn.~(\ref{eq13}), the phase constant $k_+$ can be approximated from Eqn.~(\ref{eq9}) as
\begin{eqnarray}\label{eq14}
k_{+}\simeq\sqrt{2\mu\epsilon}\omega_p\left|1-\omega/\omega_p\right|^\alpha,
\end{eqnarray}
where the opto-critical exponent yields $\alpha\rightarrow\infty$ for $\omega<\omega_p$ and $\alpha=1/2$ for $\omega>\omega_p$. Together, $k_+$ and $k_-$ dictate the opto-critical behavior of the remaining optical properties near $\omega=\omega_p$ as $\omega_p\tau\rightarrow\infty$. Using Eqn.~(\ref{eq14}), the opto-critical group velocity ($v_g=\frac{\partial\omega}{\partial k_{+}}$) becomes
\begin{eqnarray}\label{eq15}
v_g\simeq\frac{1}{\sqrt{2\mu\epsilon},\alpha}\left|1-\omega/\omega_p\right|^{1-\alpha},
\end{eqnarray}
with the exponent $1-\alpha\rightarrow-\infty$ for $\omega<\omega_p$ and $1-\alpha=1/2$ for $\omega>\omega_p$. Consequently, $v_g\rightarrow\infty$ just below $\omega_p$ and $v_g\rightarrow0$ just above $\omega_p$. The discontinuity and superluminal magnitude ($v_g > c$) of evanescent wave group velocity in a Drude metal are unsurprising, because group velocity differs from signal energy velocity, which never exceeds $c$ even in strongly dispersive media \cite{Oughstun,Zangwill}.

Using Eqns.~(\ref{eq13}) and (\ref{eq14}), the complex dielectric constant, $\tilde{\epsilon}/\epsilon_0 = [(k_+^2 - k_-^2) + i2k_+k_-]c^2/\omega^2$ (\textit{having no singularity}), simplifies to the following:
\begin{eqnarray}\label{eq16}
\tilde{\epsilon}/\epsilon_0&\simeq&2\mu\epsilon c^2\left[\omega/\omega_p-1\right]^{\beta_1}+i\frac{\mu\epsilon c^2}{\omega_{p}\tau}\left[\omega/\omega_p-1\right]^{\beta_2}\nonumber\\&&+\mathcal{O}\left(1/[\omega_p\tau]^2\right)
\end{eqnarray}
where the opto-critical exponents are $\beta_1 = 1$ for the real part and $\beta_2 = 0$ for the imaginary part, both below and above the plasma frequency. It is clear from Eqn.~(\ref{eq16}) that the dielectric constant, though small, is primarily imaginary at the opto-critical point $\omega = \omega_p$, indicating weak loss due to Ohmic conductivity even at the opto-critical point. Exponents for other optical parameters in Drude metals can, however, be derived similarly.

\section{Screening and temperature dependence in optical properties: a semi-classical theory}
Before concluding, let us note that the Drude model is a classical model. It is applicable at high temperatures ($T\gg T_F$\footnote{At such high temperatures, the Thomas-Fermi wave number ($q_0$) vanishes as $\sim1/\sqrt{T}$.}) and naturally neglects quantum (Fermi-Dirac) statistics for the conduction electrons. At low temperatures ($T\ll T_F$\footnote{Here, $T_F=\epsilon_F/k_B$ is the Fermi temperature and $\epsilon_F$ is the Fermi energy of the conduction electrons.}), conduction electrons follow Fermi-Dirac statistics \cite{Chaturvedi,Biswas-2012a} and rearrange their density to partially screen an external electric field within a characteristic length scale ($\sim\text{\AA}$), namely the Thomas-Fermi (semi-classical) screening length ($\lambda_{TF}$). The corresponding wave number ($1/\lambda_{TF}=q_0$) is called the Thomas-Fermi wave number, given by $q_0=\sqrt{\frac{e^2}{\epsilon}\frac{\partial n_e}{\partial\epsilon_F}}=\sqrt{\frac{3n_ee^2}{2\epsilon\epsilon_F}}$, where $\epsilon_F=\frac{\hbar^2}{2\vert{}m_e^*\vert{}}[3\pi^2n_e]^{2/3}$ is the Fermi energy of the conduction electrons \cite{Ashcroft}. This screening effectively renormalizes the plasma frequency from $\omega_p$ to $\bar{\omega}_p(q)=\frac{\omega_p}{\sqrt{1+[q_0/q]^2}}$ for a conduction electron with momentum $\hbar\vec{q}$ \cite{Ashcroft}. Consequently, the complex dielectric constant ($\tilde{\epsilon}/\epsilon_0=\tilde{n}^2=\frac{c^2}{\omega^2}[k_++ik_-]^2$) for a conduction electron with momentum $\hbar\vec{q}$ in the conductor is renormalized using Eqn.~(\ref{eq9}) as
\begin{eqnarray}\label{eq17}
\frac{\tilde{\epsilon}(q)}{\epsilon_0}=\frac{\epsilon/\epsilon_0}{1+\omega^2\tau^2}\bigg[1+[\omega^2-\bar{\omega}^2_p(q)]\tau^2+i\frac{\bar{\omega}^2_p(q)\tau}{\omega}\bigg]
\end{eqnarray}
for all frequencies. At low temperatures ($T\ll T_F$), the thermal average value of the complex dielectric constant ($\langle\tilde{\epsilon}\rangle/\epsilon_0$) can be obtained by replacing $\bar{\omega}^2_p(q)$ in Eqn.~(\ref{eq17}) with its Fermi-Dirac statistical average, $\langle\omega^2_p\rangle$, yielding
\begin{eqnarray}\label{eq18}
\frac{\langle\tilde{\epsilon}\rangle}{\epsilon_0}\simeq\frac{\epsilon/\epsilon_0}{1+\omega^2\tau^2}\bigg[1+[\omega^2-\langle\omega^2_p\rangle]\tau^2+i\frac{\langle\omega^2_p\rangle\tau}{\omega}\bigg],
\end{eqnarray}where the renormalized effective plasma frequency squared takes the form
\begin{eqnarray}\label{eq19}
\langle\omega^2\rangle_p&=&\frac{\int\frac{\omega^2_p}{1+[q_0/q]^2}\frac{1}{\text{e}^{[\hbar^2q^2/2|m_e^*|-\bar{\mu}]/k_BT}+1}\frac{V\text{d}^3\vec{q}}{(2\pi)^3}}{\int\frac{1}{\text{e}^{[\hbar^2q^2/2|m_e^*|-\bar{\mu}]/k_BT}+1}\frac{V\text{d}^3\vec{q}}{(2\pi)^3}}\nonumber\\&\simeq&\omega^2_p\bigg[1+\frac{3\hbar^2q_0^2}{2|m_e^*|\epsilon_F}\bigg[\frac{\hbar q_0}{\sqrt{2|m_e^*|\epsilon_F}}\cot^{-1}\big(\frac{\hbar q_0}{\sqrt{2m_e^*\epsilon_F}}\big)\nonumber\\&&-1\bigg]+\frac{\pi^2\hbar^2q_0^2|m_e^*|k_B^2T^2}{2\epsilon_F[\hbar^2q_0^2+2|m_e^*|\epsilon_F]^2}+\mathcal{O}\big(k_BT/\epsilon_F\big)^4\bigg],\nonumber\\
\end{eqnarray}
where $V$ is the volume of the conductor and $\bar{\mu}\simeq\epsilon_F[1-\pi^2k_BT^2/12\epsilon_F^2+\mathcal{O}\big([k_BT/\epsilon_F]^4\big)]$ \cite{Biswas-2012b} is the chemical potential of the conduction electrons at low temperatures ($k_BT\ll\epsilon_F$).

Eqn.~(\ref{eq18}), together with Eqn.~(\ref{eq19}), is a direct result of applying the Sommerfeld asymptotic expansion\footnote{Here, we have used the identity $\int_{-\infty}^{\infty}\frac{\rho(\epsilon)}{\text{e}^{[\epsilon-\bar{\mu}]/k_BT}+1}\text{d}\epsilon\simeq\int_{-\infty}^{\bar{\mu}}\rho(\epsilon)\text{d}\epsilon+\frac{\pi^2k_B^2T^2}{6}\rho'(\bar{\mu})+\mathcal{O}\big([k_BT/\bar{\mu}]^4\big)$, derived in Appendix C (pp. 760–761) of Ref. \cite{Ashcroft}.} to the right-hand side of Eqn.~(\ref{eq17}). This extends the description of optical properties to the Drude-Sommerfeld model, treating the conductor as a Drude-Sommerfeld metal. It is worth noting that while the second term in Eqn.~(\ref{eq19}) accounts purely for the quantum many-body screening effect on the plasma frequency, the third term provides its low-temperature thermal correction. Whereas quantum many-body screening reduces the renormalized plasma frequency squared ($\langle\omega^2_p\rangle$) from $\langle\omega^2_p\rangle=\omega^2_p$ at $q_0=0$ down to $\langle\omega^2_p\rangle=0$ as $q_0\rightarrow\infty$, the thermal contribution increases it as the screening wave number increases. In the high-temperature limit ($T\rightarrow\infty$), we recover $\langle\omega^2_p\rangle=\omega^2_p$ because screening vanishes ($q_0\rightarrow0$). Thus, $q_0=0$ represents the classical limit of the semi-classical (and full quantum-mechanical) theory. This naturally leads to no corrections to either the complex dielectric constant or any associated optical properties (including the attenuation constant, which can be derived directly from the complex dielectric constant). Consequently, quantum corrections to the optical properties of a Drude-Sommerfeld metal leave its main physical features unchanged, serving primarily to reduce the effective plasma frequency ($\langle\omega_p\rangle\simeq\sqrt{\langle\omega^2_p\rangle}$).

\section{Conclusion}
To conclude, we have analytically determined the attenuation constant [Eqn.~$(\ref{eq9})$, Fig.~\ref{fig1}], phase constant [Eqn.~$(\ref{eq9})$], and reflectivity [Eqn.~$(\ref{eq13a})$] of Drude metals with background bound charges and currents across all frequencies ($0<\omega<\infty$), thereby unifying existing low- and high-frequency results in a relatively straightforward manner. Our results [Eqn.~\ref{eq13a}, Fig.~\ref{fig2}] have shown good agreement with existing experimental data on the reflectivity of dense Z-pinch plasmas in air and hydrogen (good conductors) \cite{Kurilenkov,Skowronek} and seawater (a relatively poor conductor) \cite{Saxton}. Focusing on high carrier concentrations ($\omega_p\tau\gg1$), we have obtained a simple expression for the attenuation constant, $k_-\simeq+\sqrt{\frac{\mu\epsilon}{2}}\sqrt{\omega_p^2-\omega^2+\vert{}\omega_p^2-\omega^2\vert{}}$ [Eqn.~$(\ref{eq12})$], which is valid near the plasma frequency $\omega_p$ but possesses a branch-point singularity at $\omega=\omega_p$. Consequently, near $\omega_p$, the optical properties—including the attenuation constant, phase constant, group velocity, and complex dielectric constant—have exhibited opto-critical behavior [Eqns.~$(\ref{eq14})$–$(\ref{eq16})$], for which we have calculated the opto-critical exponents in analogy with second-order phase transitions in statistical mechanics. To address thermal effects below the Fermi temperature ($T\ll T_F$), where the classical Drude model fails, we have applied the Drude–Sommerfeld model with Thomas–Fermi screening to incorporate quantum corrections to the plasma frequency [Eqns.~$(\ref{eq18})$ and $(\ref{eq19})$]. Our work may find applications in optics, condensed matter physics, and physical oceanography—even for transmitting high-frequency signals ($\sim 500\text{ GHz}$) near the plasma frequency to submarines to achieve low reflectivity and high transmissivity.

The opto-criticality we have explored is not the criticality often discussed in connection with continuous phase transitions in statistical mechanics \cite{Wilson,Fisher} or dynamic criticality \cite{Halperin,Ferrell,Bhattacharjee}. It is, rather, the criticality of optical properties in the frequency domain around the plasma frequency.

Here, we haven't considered frequency-independent permittivity and permeability ($\epsilon$ and $\mu$, respectively) from bound charges and currents in the background conductor when solving Eqns.~(\ref{eq1}). We have made this approximation by considering the medium to be a linear one. This assumption holds over long time scales ($t\gg\epsilon/\sigma_0$) and for all non-zero frequencies ($\omega>0$). Static equilibrium, however, is not achieved for oscillatory fields ($\omega>0$). Under static conditions ($\omega=0$ and $t\gg\epsilon/\sigma_0$) without an applied voltage, exposure to an electrostatic field causes relative permittivity $\epsilon/\epsilon_0$ to approach a high value ($\gtrsim1$) for good conductors, and infinity for perfect conductors ($\sigma_0\rightarrow\infty$). We have not considered this static case, as optical properties strictly require $\omega>0$. In the quasi-static regime ($\omega\tau\ll1$), permittivity and permeability are frequency-independent. In the non-quasi-static regime—especially at high frequencies—they depend weakly on frequency (compared to the strong frequency dependence of the conductor’s optical response) and approach free-space values $\epsilon_0$ and $\mu_0$ as $\omega\tau\rightarrow\infty$ \cite{Zangwill}. We have left the investigation of this weak frequency dependence as an open problem.

Figure \ref{fig2} shows that our result [Eqn.~(\ref{eq13a})] matches with experimental data better for higher $\omega_{p}\tau$ (plasmas \cite{Kurilenkov}) than for lower $\omega_{p}\tau$ (seawater \cite{Saxton}). This discrepancy may reflect a limitation of the Drude or Drude–Sommerfeld models. Alternatively, it may arise from experimental parameters, specifically the relaxation time and static dielectric constant \cite{Saxton}, which were measured for an NaCl solution rather than seawater itself.

The temperature dependence of optical properties has been primarily explored at low temperatures ($T\ll T_D\ll T_F$, where $T_D$ is the Debye temperature). Here, the scattering rate ($\tau^{-1}$) is assumed to be temperature-independent, mainly due to inelastic electron–impurity collisions. However, substantial temperature dependence also arises at low temperatures from electron–electron ($\tau^{-1}_{\text{el-el}}\propto T^2$) and electron–phonon interactions ($\tau^{-1}_{\text{el-ph}}\propto T^5$) \cite{Annett}. At high temperatures ($T\gg T_D$), electron–phonon interactions dominate, yielding $\tau^{-1}_{\text{el-ph}}\propto T$ \cite{Tu}. Although the classical Drude model applies at high temperatures, we apply it considering only electron–impurity scattering to compare Drude and Drude–Sommerfeld metals at low temperatures \cite{Ashcroft2}. Conduction electrons, however, are to be replaced with charge carriers wherever applicable, e.g., ions in seawater. Generalization of our semi-classical theory with the Drude-Sommerfeld model, by including the other two scattering effects in both the low-temperature and high-temperature regimes, is kept as an open problem.

We have considered only the normal incidence of the plane electromagnetic wave for the study of the optical properties of the Drude and Drude-Sommerfeld metals. The surface plasmons \cite{Otto,Kretschmann}, however, come into the picture exclusively for the TM-polarized (p-polarized) modes at a metal-dielectric interface for an oblique incidence \cite{Biswas2022} especially for the high-frequency regime \cite{Pendry2015,Garcia-Vidal2010,Saxton}. The study of the opto-criticality for the oblique incidence, as a generalization of our work, is kept as an open problem.

Thomas–Fermi screening within the Drude–Sommerfeld model does not fundamentally alter the optical properties predicted by the Drude model, except by modifying the effective plasma frequency.

Finally, we have used the Thomas–Fermi screening for convenience. It is actually a special case of the Lindhard screening \cite{Ashcroft}, which becomes relevant at long distances when the wave-number ($q$) of a conduction electron is much smaller than the Fermi wave-number ($q_F$). The study of the optical properties with Lindhard screening has also been left as an open problem.

\section*{Acknowledgement}
S. Biswas acknowledges the partial financial support of the SERB (now ANRF), DST, Govt. of India under the EMEQ Scheme [No. EEQ/2023/000788]. Useful discussions with Dr. Abhiram Soori (UoH, India) and Prof. N. Sri Ram Gopal (UoH, India) are gratefully acknowledged.

\section*{Author Contributions}
B. K. Behera solved the problem and wrote the article. R. Bhattacharyya solved the problem and wrote the article. S. Biswas framed the problem, solved the problem, and wrote the article.

\subsection*{Data Availability Statement} 
No new data were created or analyzed in this study.

\subsection*{Declarations}
\textbf{Conflict of Interest}: The authors declare no conflicts of interest.


\begin{thebibliography}{99}
\bibitem{Griffiths}D. J. Griffiths, \textit{Introduction to Electrodynamics}, 4th edn., Sec. 9.4.1, pp. 412-415 and Sec. 7.3.5, pp. 340-340 (Pearson, Boston, 2013)

\bibitem{Born}M. Born and E. Wolf, \textit{Principles of Optics}, 7th edn., Ch. 14, pp. 735-789 (Cambridge University Press, Cambridge, 1999) 

\bibitem{Roberts}S. Roberts, \textit{Optical properties of copper}, \href{https://doi.org/10.1103/PhysRev.118.1509}{\prv{118}{1509}{1960}}

\bibitem{Linz}A. Linz, Jr. and R. E. Newnham, \textit{Ultraviolet absorption spectra in ruby}, \href{https://doi.org/10.1103/PhysRev.123.500}{\prv{123}{500}{1961}}

\bibitem{Wooten}F. Wooten, \textit{Optical Properties of Solids}, Ch. 2-4, pp. 15-84 (Academic Press, New York, 1972)

\bibitem{Fox}M. Fox, \textit{Optical Properties of Solids}, Ch. 2, pp. 25-84 \& Ch. 7. pp. 143-164  (Oxford University Press, Oxford, 2001)

\bibitem{Yarmohammadi}M. Yarmohammadi, M. Bukov, and M. H. Kolodrubetz, \textit{Nonequilibrium phononic first-order phase transition in a driven fermion chain}, \href{https://doi.org/10.1103/PhysRevB.108.L140305}{\prb{108}{L140305}{2023}}

\bibitem{Murakami}Y. Murakami, D. Golež, M. Eckstein, and P. Werner, \textit{Photoinduced nonequilibrium states in Mott insulators}, \href{https://doi.org/10.1103/tkjh-lr83}{\rmp{97}{035001}{2025}}

\bibitem{Cardona}M. Cardona and M. L. W. Thewalt, \textit{Isotope effects on the optical spectra of semiconductors}, \href{https://doi.org/10.1103/RevModPhys.77.1173}{\rmp{77}{1173}{2005}}

\bibitem{Devereaux}T. P. Devereaux and R. Hackl, \textit{Inelastic light scattering from correlated electrons}, \href{https://doi.org/10.1103/RevModPhys.79.175}{\rmp{79}{175}{2007}}

\bibitem{Blaber}M. G. Blaber, M. D. Arnold, and M. J. Ford, \textit{A review of the optical properties of alloys and intermetallics for plasmonics}, \href{https://doi.org/10.1088/0953-8984/22/14/143201}{J. Phys.: Condens. Matter \textbf{22} 143201 (2010)}

\bibitem{Guistino}F. Giustino, \textit{Electron-phonon interactions from first principles}, \href{https://doi.org/10.1103/RevModPhys.89.015003}{\rmp{89}{015003}{2017}}

\bibitem{Russo}E. D. Russo, T. Verstijnen, P. Koenraad, K. Pantzas, G. Patriarche, and L. Rigutti, \textit{Order and disorder at the atomic scale: Microscopy applied to semiconductors}, \href{https://doi.org/10.1103/RevModPhys.97.025006}{\rmp{97}{025006}{2025}}

\bibitem{Vries}N. de Vries, J. Chen, E. Hoglund, X. Guo, D. Chaudhuri, J. Hachtel, and P. Abbamonte, \textit{Comparative analysis of plasmon modes in layered Lindhard metals and strange metals}, \href{https://doi.org/10.1103/39wl-mzly}{\prb{112}{165145}{2025}}

\bibitem{Arafune}R. Arafune, H. Ishida, C.-L. Lin, and N. Takagi, \textit{Probing moiré Bloch bands of photoexcited electrons on graphene/Ir(111)}, \href{https://doi.org/10.1103/yg4g-x51r}{\prb{112}{L161408}{2025}}

\bibitem{Tonauer}C. M. Tonauer, E.-M. Köck, R. Henn, C. Kappacher, C. W. Huck, and T. Loerting, \textit{Near-Infrared spectroscopic sensing of hydrogen order in ice XIII}, \href{https://doi.org/10.1103/x2ph-yp2v}{\prl{135}{018002}{2025}}

\bibitem{Rahaman}I. Rahaman, A. Bolda, B. Li, H. D. Ellis, and K. Fu, \textit{Study of optical properties of MOCVD-grown rutile GeO2 films}, \href{https://doi.org/10.1088/1361-6463/ae13ba}{J. Phys. D: Appl. Phys. \textbf{58}, 43LT01 (2025)}

\bibitem{Tiwari}V. Tiwari, R. Korol, and I. Franco, \textit{Robust purely optical signatures of Floquet states in laser-dressed crystals}, \href{https://doi.org/10.1103/5ywx-7dbs}{\prl{135}{186901}{2025}}

\bibitem{Huang}T. Huang and Z. Sun, \textit{Universal phase transitions of matter in optically driven cavities}, \href{https://doi.org/10.1103/71fd-m6rj}{\prl{136}{036901}{2026}}

\bibitem{Talley}L. D. Talley, G. L. Pickard, W. J. Emery, and J. H. Swift, \textit{Descriptive Physical Oceanography: An Introduction}, 6th edn. (Elsevier, Amsterdam, 2011)  

\bibitem{Adcock}T. A. A. Adcock and P. H. Taylor, \textit{The physics of anomalous (`rogue’) ocean waves}, \href{https://doi.org/10.1088/0034-4885/77/10/105901}{Rep. Prog. Phys. \textbf{77}, 105901 (2014)}

\bibitem{Jarman}S. Jarman, \textit{Ice loss is transforming the light-absorption properties of seawater}, \href{http://link.aps.org/doi/10.1103/Physics.18.109}{Physics \textbf{18}, 109 (2025)}

\bibitem{Moore}R. K. Moore, \textit{Radio communication in the sea}, \href{https://doi.org/10.1109/MSPEC.1967.5217169}{IEEE Spectrum \textbf{4}, 42 (1967)}

\bibitem{Smolyaninov}I. I. Smolyaninov, Q. Balzano, M. Barry, and D. Young, \textit{Superlensing enables radio communication and imaging underwater}, \href{https://doi.org/10.1038/s41598-023-45663-6}{Sci. Rep. \textbf{13}, 18333 (2023)}

\bibitem{Rothwell}E. J. Rothwell and M. J. Cloud, \textit{Electromagnetics}, edn. 3, Sec. 3.6.2.3, pp. 253-255, CRC Press (Taylor and Francis Group), Boca Raton (2018)

\bibitem{Drude}P. Drude, \textit{Zur elektronentheorie der metalle}, \href{https://doi.org/10.1002/andp.19003060312}{Ann. der Physik \textbf{306}, 566 (1900)} \& \textit{Zur elektronentheorie der metalle; II. Teil. Galvanomagnetische und thermomagnetische effecte}, \href{https://doi.org/10.1002/andp.19003081102}{Ann. der Physik \textbf{308}, 369 (1900)}

\bibitem{Jackson}J. D. Jackson, \textit{Classical Electrodynamics}, 3rd edn., Sec. 7.5, pp. 309-316, (Wiley, New York, 1999)

\bibitem{Zangwill}A. Zangwill, \textit{Modern Electrodynamics}, Sec. 14.7 pp. 472-481 \& Sec. 18.5-6, pp. 630-648, Cambridge University Press, Cambridge (2013)

\bibitem{Biswas-2026}S. Biswas, \textit{Gauss' law for arbitrarily moving Gaussian surfaces: a revisit}, \href{https://doi.org/10.1007/s12045-026-2059-3}{Reson. \textbf{31}, 1313 (2026)}

\bibitem{Bronin}S. Ya. Bronin, B. B. Zelener, and B. V. Zelener, \textit{Refraction, absorption and reflectivity of radiation in strongly coupled plasma}, \href{https://doi.org/10.1016/j.jqsrt.2021.107621}{J. Quant. Spectrosc. Radiat. Transf. \textbf{268}, 107621 (2021)}

\bibitem{Oughstun}K. E. Oughstun, \textit{Electromagnetic and Optical Pulse Propagation}, Vol. 1: \textit{Spectral Representations in Temporally Dispersive Media Second Edition}, edn. 2, Sec. 5.2.6, pp. 300-306 (Springer Nature, Cham, 2019)

\bibitem{Boschini}F. Boschini, M. Zonno and A. Damascelli, \textit{Time-resolved ARPES studies of quantum materials}, \href{https://doi.org/10.1103/RevModPhys.96.015003}{\rmp{96}{015003}{2024}}

\bibitem{WilsonNL}K. G. Wilson, \textit{The renormalization group and critical phenomena}, \href{https://doi.org/10.1103/RevModPhys.55.583}{\rmp{55}{583}{1983}}

\bibitem{Kurilenkov}Yu. K. Kurilenkov and M. A. Berkovsky, \textit{Collective Modes and Correlations}, In: G. A. Kobzev, I. T. Iakubov, and M. M. Popovich (eds), \textit{Transport and Optical Properties of Nonideal Plasma}, Sec. 6.2.1, pp. 272–274 (Springer Science+Business Media, New York, 1995)

\bibitem{Saxton}J. A. Saxton and J. A. Lane, \textit{Electrical properties of sea water: reflection and attenuation characteristics at V.H.F.}, Wireless Engineer: The Journal of Radio Research and Progress (World Radio History) \textbf{29}, 269 (1952)

\bibitem{Sommerfeld}A. Sommerfeld, \textit{Zur Elektronentheorie der Metalle auf Grund der Fermischen Statistik}, \href{https://doi.org/10.1007/BF01391052}{Z. für Physik \textbf{47}, 1 (1928)}

\bibitem{Thomas-Fermi}L. H. Thomas, \textit{The calculation of atomic fields}, \href{https://doi.org/10.1017/S0305004100011683}{Math. Proc. Cam. Philos. Soc. \textbf{23}, 542 (1927)}; E. Fermi, \textit{Eine statistische methode zur bestimmung einiger eigenschaften des atoms und ihre anwendung auf die theorie des periodischen systems der elemente}, \href{https://doi.org/10.1007/BF01351576}{Z. für Physik \textbf{48}, 73 (1928)}

\bibitem{Skowronek}M. Skowronek, J. Rous, A. Goldstein, and F. Cabannes, \textit{Influence of plasma frequency on the light emitted by an exploding ionized gaseous filament}, \href{https://doi.org/10.1063/1.1692929}{Phys. Fluids (1958-1988) \textbf{13}, 378 (1970)}

\bibitem{Reinholz}H. Reinholz, Yu. Zaporoghets, V. Mintsev, and V. Fortov, \textit{Frequency-dependent reflectivity of shock-compressed xenon plasmas}, \href{https://doi.org/10.1103/PhysRevE.68.036403}{\pre{68}{036403}{2003}}
 
\bibitem{Verleur}H. W. Verleur, \textit{Determination of optical constants from reflectance or transmittance measurements on bulk crystals or thin films}, \href{https://doi.org/10.1364/JOSA.58.001356}{J. Opt. Soc. Am. \textbf{58}, 1356 (1968)}

\bibitem{Behera}B. K. Behera, S. K. Gour, and S. Biswas, \textit{Refractive index for the mechanical refraction of a relativistic particle}, \href{https://doi.org/10.1140/epjd/s10053-024-00849-z}{\epjd{78}{60}{2024}}

\bibitem{Voronin}A. A. Voronin and A. M. Zheltikov, \textit{The generalized Sellmeier equation for air}, \href{https://doi.org/10.1038/srep46111}{Sci. Rep. \textbf{7}, 46111 (2017)}

\bibitem{Harvey}A. H. Harvey, J. S. Gallagher, and J. M. H. Levelt Sengers, \textit{Revised formulation for the refractive index of water and steam as a function of wavelength, temperature and density}, \href{https://doi.org/10.1063/1.556029}{J. Phys. Chem. Ref. Data \textbf{27}, 761 (1998)}

\bibitem{Chaturvedi}S. Chaturvedi and S. Biswas, \textit{Fermi-Dirac statistics}, \href{https://doi.org/10.1007/s12045-014-0006-1}{Reson. \textbf{19}, 45 (2014)}

\bibitem{Biswas-2012a}S. Biswas, D. Jana, and R. K. Manna, \textit{Excess energy of an ultracold Fermi gas in a trapped geometry}, \href{https://doi.org/10.1140/epjd/e2012-30152-y}{\epjd{66}{217}{2012}}

\bibitem{Ashcroft}N. W. Ashcroft and N. D. Mermin, \textit{Solid State Physics}, Ch. 17, pp. 340-344 (Harcourt Asia, Singapore, 1976)

\bibitem{Biswas-2012b}S. Biswas and D. Jana, \textit{Thermodynamics of quantum gases for the entire range of temperature}, \href{https://doi.org/10.1088/0143-0807/33/6/1527}{{\ejp{33}{1527}{2012}}}

\bibitem{Wilson}K. G. Wilson, \textit{Renormalization group and critical phenomena. I. Renormalization group and the Kadanoff scaling picture}, \href{https://doi.org/10.1103/PhysRevB.4.3174}{\prb{4}{3174}{1971}} 

\bibitem{Fisher}M. E. Fisher, \textit{The renormalization group in the theory of critical behavior}, \href{https://doi.org/10.1103/RevModPhys.46.597}{\rmp{46}{597}{1974}}; Erratum: \href{https://doi.org/10.1103/RevModPhys.47.543}{Rev. Mod. Phys. \textbf{47}, 543 (1975)}

\bibitem{Halperin}B. I. Halperin, P. C. Hohenberg, and S.-K. Ma, \textit{Calculation of dynamic critical properties using Wilson's expansion methods}, \href{https://doi.org/10.1103/PhysRevLett.29.1548}{\prl{29}{1548}{1972}}

\bibitem{Ferrell}R. A. Ferrell and J. K. Bhattacharjee, \textit{Corrections to dynamic scaling for the lambda transition in liquid helium}, \href{https://doi.org/10.1103/PhysRevLett.42.1638}{\prl{42}{1638}{1979}}

\bibitem{Bhattacharjee}J. K. Bhattacharjee and R. A. Ferrell, \textit{Frequency-dependent critical diffusion in a classical fluid}, \href{https://doi.org/10.1103/PhysRevA.23.1511}{\pra{23}{1511}{1981}}

\bibitem{Annett}J. F. Annett, \textit{Superconductivity, Superfluids and Condensates}, Sec. 3.2, pp. 47-49 (Oxford University Press, Oxford, 2004)

\bibitem{Tu}Y.-T. Tu and S. D. Sarma, \textit{Negative intercept of the apparent zero-temperature extrapolated linear-in-T metallic resistivity}, \href{https://doi.org/10.1103/PhysRevB.110.075151}{\prb{110}{075151}{2024}}

\bibitem{Ashcroft2}See Ch. 1, pp. 20-25 of Ref.\cite{Ashcroft}.

\bibitem{Otto}A. Otto, \textit{Excitation of nonradiative surface plasma waves in silver by the method of frustrated total reflection}, \href{https://doi.org/10.1007/BF01391532}{Z. Physik \textbf{216}, 398 (1968)}

\bibitem{Kretschmann}E. Kretschmann and H. Raether, \textit{Notizen: Radiative decay of non radiative surface plasmons excited by light}, \href{https://doi.org/10.1515/zna-1968-1247}{Z. Naturforsch.  A \textbf{23}, 2135 (1968)}

\bibitem{Biswas2022}S. Biswas, R. Bhattacharyya, and S. Bhattacharyya, \textit{Explicit derivation of the Fraunhofer diffraction formula for oblique incidence}, \href{https://doi.org/10.1088/1361-6404/ac2ece}{\ejp{43}{015301}{2022}}

\bibitem{Pendry2015}J. B. Pendry, Y. Luo, and R. Zaho, \textit{Transforming the optical landscape}, \href{https://doi.org/10.1126/science.1261244}{\sc{348}{521}{2015}}

\bibitem{Garcia-Vidal2010}F. J. Garcia-Vidal, L. Martin-Moreno, T. W. Ebbesen, and L. Kuipers, \textit{Light passing through subwavelength apertures}, \href{https://doi.org/10.1103/RevModPhys.82.729}{\rmp{82}{729}{2010}}
\end{thebibliography}
\end{document}